# Nuclei Segmentation in Histopathology Images using Deep Learning with Local and Global Views

Mahdi Arab Loodaricheh[1], Nader Karimi[1], Shadrokh Samavi[1,2]
[1]Dept. of Elect. and Comp. Eng., Isfahan University of Technology, Isfahan, 84156-83111 Iran
[2]Dept. of Elect. and Comp. Eng., McMaster University, Hamilton, ON L8S 4L8, Canada

*Abstract*—*Digital pathology is one of the most significant developments in modern medicine. Pathological examinations are the gold standard of medical protocols and play a fundamental role in diagnosis. Recently, with the advent of digital scanners, tissue histopathology slides can now be digitized and stored as digital images. As a result, digitized histopathological tissues can be used in computer-aided image analysis programs and machine learning techniques. Detection and segmentation of nuclei are some of the essential steps in the diagnosis of cancers. Recently, deep learning has been used for nuclei segmentation. However, one of the problems in deep learning methods for nuclei segmentation is the lack of information from out of the patches. This paper proposes a deep learning-based approach for nuclei segmentation, which addresses the problem of misprediction in patch border areas. We use both local and global patches to predict the final segmentation map. Experimental results on the Multi-organ histopathology dataset demonstrate that our method outperforms the baseline nuclei segmentation and popular segmentation models.*

## I. INTRODUCTION

Histopathology has an essential role in cancer diagnosis. Diagnosis of the disease after manual analysis of biopsy slides is a laborious task for pathologists. With the advancement of digital pathology, the automatic detection of patterns in whole slide images will help pathologists for their daily diagnoses [1]. Histopathological images have biologically normal and abnormal structures, morphological features, and shapes that pathologists can identify depending on their experience. However, some features, such as tissue areas and cell patterns, have visual diversity and usually are hidden from the pathologist. The visual variety of these images stems from the biological and anatomical structures of the tissue [2]. In traditional methods of cancer diagnosis, the pathologist analyzes tissue biopsy and makes a diagnosis based on the presence or absence of cancer depending on the morphology and structure of the cells. But in recent years, computer methods in pathology have been responsible for nuclei segmentation in histopathological imaging and cancer diagnosis [3]. Computer-aided diagnosis (CAD) significantly increases the efficiency and accuracy of pathologist decisions and can generally be beneficial to the patient. In research fields, segmenting the nucleus of cells and classifying them is a repetitive task, which is very difficult in pathology images. Detecting and segmenting the nucleus in cytopathological images is easier because of the isolated nuclei and the absence of complex tissue structures. However, researchers are trying to continue to diagnose cancer and segment images in histopathology images and increase accuracy and efficiency in this area [4].

Nuclei detection or detecting the location of nuclei in histopathological images is a task in which the boundaries of the nuclei are not precisely separated. The reason for doing this task is to approximate the location of the nuclei and use these markers for the operation of counting the nuclei, nuclei tracking, or the final segmentation by these markers. In contrast, nuclei segmentation is more complex than nuclei detection. Because borders should be determined relatively accurately; however, in the nuclei detection, the location of the nucleus and a specific marker of this part is sufficient for pathologists in some tasks, and there is no need for the actual borders [5]. Having histopathology nuclei and analyzing their morphology and texture will help the pathologist diagnose the disease's type and degree. In addition, the quality of this diagnosis and classification is directly related to the quality of histopathological image segmentation. Nuclei segmentation methods can be divided into traditional methods and learning-based methods. Traditional methods include using thresholding [6], watershed transform [7], and clustering [8]. Learning-based methods consist of traditional supervised learning methods like SVM and deep neural networks. While deep learning methods need a large number of learning data, the performance of these methods is generally superior to other learning methods. Fully convolutional neural networks are the most common architectures in image segmentation. One of the most important architectures used in medical image segmentation is U-Net architecture. U-Net [9] is an FCN based network architecture with an encoder part for downsampling and a decoder part for upsampling. In addition, skip-connection, which was introduced in residual networks, is applied to the U-Net architecture. The current deep learning method for nuclei segmentation mostly needs a post-processing algorithm, and they have no information about the global view. Naylor [10] uses FCN architecture and then applies a watershed algorithm in the post-processing phase. In another work, Naylor [11] proposed a method for nuclei segmentation by formulating the segmentation problem as a regression task of the distance map. They use the watershed algorithm for post-processing. Kumar [12] proposed a CNN3 model to predict nuclei and their boundary in the image. Moreover, Zheng [13] proposed RIC-

U-net, which uses residual blocks, multi-scale, and channel attention mechanism; however, this method has a problem with complicated cases with unclear pixel differences between foreground and background. However, none of these methods use global information available in whole slide images.

This paper proposed an end-to-end method based on fully convolutional neural network architecture, a U-Net backbone, and an EfficientNet encoder. We used local and global patches to predict the nuclei segmentation map more accurately. Using global patches can address the problem of misprediction in patches border areas. Moreover, we used stain normalization and global patch extraction in the preprocessing phase.

The rest of the paper is arranged as follows. The proposed method is reported in section II. The implementation of our architecture, along with results, is included in section III. The conclusion is drawn in section IV.

## II. METHODS

Our segmentation model contains a preprocessing phase and a model architecture to predict the final segmentation result. Fig. 1 shows the proposed preprocessing diagram, and Fig. 2 shows our model architecture parts. In the following, the details of these parts are explained.

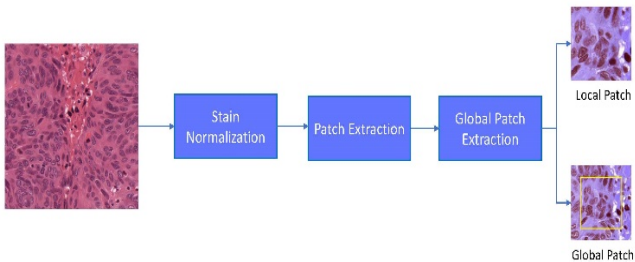

Fig. 1. Overview of our preprocessing phase

### A. Stain Normalization

Tissue images are mostly stained by the H&E staining protocol. The nuclei are colored blue in this method, and the cytoplasm is stained pink. However, this method can result in notable color variation due to differences in staining protocols and scanners; therefore, this variation can make some differences between training and testing images. Normalizing histopathology images will cause a remarkable improvement in the performance of machine learning segmentation methods. There are several methods for color normalization in histopathology images [14]–[16]. We tested three approaches proposed in [14], [15], and [16] to normalize images. For our segmentation framework, we got better results by using the method proposed by Macenko [16] to normalize images. In this method, we should choose a target image, which will convert all other images' color space to the target image's color space. The overview of this method is shown in Fig. 3.

### B. Patch Extraction

Nowadays, U-Net is one of the most important architecture in medical image segmentation. However, due to limited GPU's memory, it is not feasible to feed whole-slide histopathology images to the U-Net architecture. As a result, these images should be cut into sub-patches. We extracted 256x256 patches from whole slide images with the size of 1000x1000.

### C. Global Patch Extraction

The proposed architecture uses two inputs for the learning process. Our proposed network predicts a segmentation map based on each patch's global and local view. By using global patches in input, our focus is to solve the lack of information in the border part of the patches. Therefore, we generate a global patch that is extended 64 pixels in every direction for each patch extracted from the original image. As a result, the original patch is in the middle part of the global patch. We use zero-padding for each dimension for border parts of 1000x1000 whole slide images. The tissue patch sample and its global patch are shown in Fig. 4.

### D. Model Architecture

Our model architecture is a U-Net backbone network containing EfficientNet [17] encoder and a U-Net decoder illustrated in Fig. 2.

*1) EfficientNet Encoder:* Scaling convolutional neural networks is a process that is hard to understand and has a lot of parameters to set. However, the EfficientNet [17] is a recently proposed architecture and a method to uniformly scale the dimensions of width, depth, and resolution. This method shows the importance of balanced dimensions of width, depth, and resolution to improve the accuracy and efficiency of a model. This scaling method was used to generate a family of EfficientNet architectures containing B0 to B7. In our network architecture, we use the EfficientNet-B4 in the encoder part. Local and global patches are concatenated in the first layer of the encoder part. As a result, the global information for each patch can be propagated in the network.

*2) U-Net Decoder:* Our model decoder block has to predict nuclei segmentation map. U-Net decoder architecture is used in our method. This part is connected to the encoder through a sequence of residual blocks. The last layer of our decoder is connected to the global and local patch by residual blocks. Consequently, the prediction and the network's loss are based on a local view and global view of a tissue image, which can solve the problem of misprediction in the patches' borders. It can be inferred that the network uses the broader nuclei area's available information to optimize its weights.

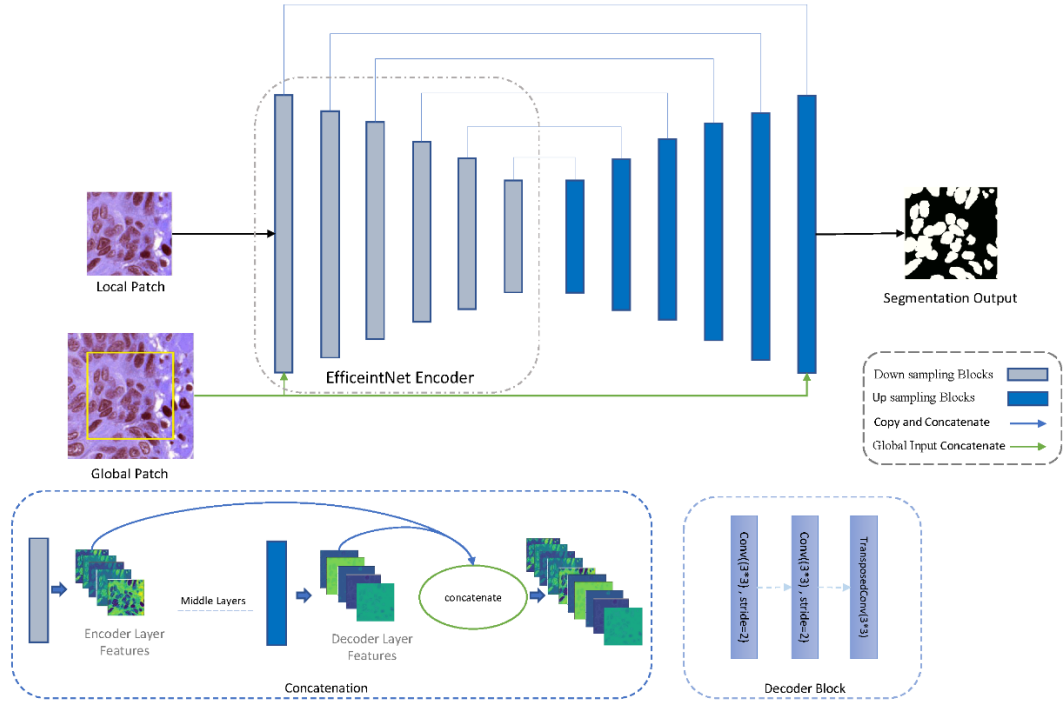

*Fig. 2 Our segmentation model architecture*

*E. Model's Loss*

Our model's loss function is based on the loss of the decoder's parts. Nuclei segmentation map prediction is a binary classification task of each pixel in the image. Hence, we used Jaccard Distance loss for segmentation. The Jaccard Distance loss function is defined by:

$$d_j(Y, Y^*) = \frac{|Y \cup Y^*| - |Y \cap Y^*|}{|Y \cup Y^*|} \quad (1)$$

where $Y$ represents the label image's pixels and $Y^*$ represents the predicted image's pixels.

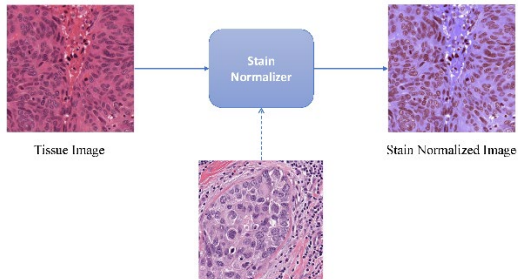

*Fig. 3. Overview of our stain normalization*

## III. EXPERIMENTS AND RESULTS

In the following, we introduce the dataset we used in this research and present details of our model implementation and its results compared to state-of-the-art methods.

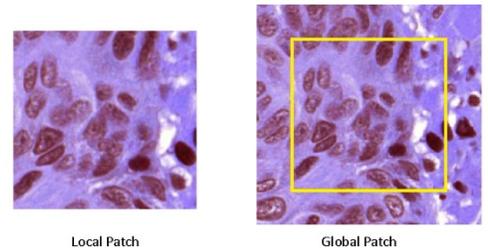

*Fig. 4. Sample of local patch and its global patch*

*A. Dataset*

We used a publicly available multi-organ pathology dataset that was thoroughly annotated and presented by Kumar [12]. This includes 30 whole-slide histopathology images from seven organs (liver, breast, prostate, kidney, colon, stomach, and bladder). The size of each image in this dataset is 1000x1000, and this dataset is from The Cancer Genome Atlas (TCGA) at ×40 magnification. We used 16 images for training and 14 images for testing in this experiment. We used the same train and test split which was used in [12].

*B. Evaluation Criteria*

The pixel-level evaluation was used to measure the performance of our nuclei segmentation method. The common method to evaluate segmentation is Dice Coefficient, which is defined as:

$$Dice = \frac{2|G \cap S|}{|G|+|S|} \quad (2)$$

where |G| and |S| represent the ground truth image's pixels and segmented image's pixels, respectively.

*C. Implementation Details*

We implemented our model using the open-source library Keras, Tensorflow, and performed it on a computer with GPU GTX 1080TI. We used several data augmentation methods such as flipping and the rotation of 90° and 180° to avoid over-fitting on the training set. We initialize the parameters of the network's Encoder part from the architecture, which is pre-trained on Imagenet. Adam optimizer variant called AMSGrad was used for the training process, and the learning rate was initially set to 0.001. We set the mini-batch size to 8 and the number of epochs set to 50. We also used 0.5 as the threshold.

Table I shows the result for the test set compared to other nuclei segmentation methods. These methods include U-Net [9], Mask R-CNN [18], CNN3 [12], and DIST [11]. Our method outperforms these models based on Dice Score. Moreover, from Fig.5, we can see some histopathology patches and their ground truth and our model's segmentation result compared to U-Net's segmentation results. We can see that our segmentation model has a superior performance and rarely predicts incorrectly in border areas.

TABLE I. COMPARATIVE ANALYSIS OF SEGMENTATION PERFORMANCE ON MONUSEG

| Method | Dice Coefficient |
| --- | --- |
| U-Net [9] | 0.758 |
| Mask-RCNN [18] | 0.760 |
| CNN3 [12] | 0.762 |
| DIST [11] | 0.789 |
| **Our Method** | **0.814** |

IV. CONCLUSION

Due to limited GPU memory, deep learning methods for the segmentation of histopathology images cannot be applied to the whole slide images without the patch extraction method. However, each patch does not have information about its covered area. The proposed framework segments each patch of the whole slide histopathology image using global patch information available in the dataset. We prepare our stain normalized local and global patches of whole slide images in the first stage. In the second stage, we trained our model which its architecture is based on U-Net, and the encoder part is an EfficientNet encoder. Our result is significantly better than baseline methods and common architectures based on Dice Score. For future works, we can extend the same procedure for multiple abnormality detection [19], brain tumor segmentation [20], vessel segmentation in angiograms [21], and segmentation of abdominal regions in CT images [22].

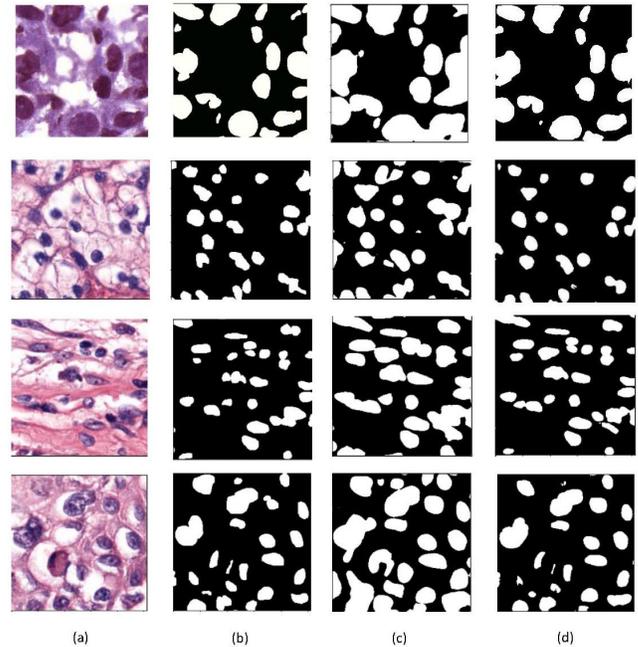

Fig. 5. Four patches results. (a) Patch of whole slide image (b) Patch GT. (c) U-Net Segmentation (d) Our model Segmentation